\documentclass[twocolumn,secnumarabic,amssymb, nobibnotes, aps, prd]{revtex4-2}

\usepackage{amsmath} 
\usepackage{graphicx}
\usepackage{epstopdf} 
\usepackage{float} 

\begin{document}

\title{Multiple phases in $\mathbf{K_2Cr_3As_3}$: a playground for manipulating topological superconductivity}%

\author{$\mathrm{Seigo~Ogawa^1}$}
\author{$\mathrm{Tomoki~Miyoshi^1}$}
\author{$\mathrm{Saki~Uchida^1}$}
\author{$\mathrm{Kazuaki~Matano^{1,\dagger}}$}
\author{$\mathrm{Shinji~Kawasaki^1}$}
\author{$\mathrm{Yoshihiko~Inada^2}$}
\author{Guo-qing~$\mathrm{Zheng^1}$}%
\email[Contact author: ]{zheng@psun.phys.okayama-u.ac.jp}
\affiliation{$\mathrm{^1}$Department of Physics, Okayama University, Okayama 700-8530, Japan}
\affiliation{$\mathrm{^2}$Faculty of Education, Okayama University, Okayama, 700-8530, Japan}
\affiliation{$\mathrm{^\dagger}$Present address: Department of Physics, Okayama University of Science, Okayama, 700-0005, Japan.}
\date{\today}%

\begin{abstract}
Spin-triplet topological superconductors are rare but of fundamental interest as they can host Majorana bound states that can be used in fault-tolerant quantum computing. Recent efforts have been devoted to searching for spin-triplet states in U-based compounds, but these materials have a low transition temperature ($T_{\mathrm{c}}$) and coexisting competing orders, which creates significant experimental challenges and often leads to contradictory conclusions. The Cr-based candidate $\mathrm{K_2Cr_3As_3}$ offers a promising alternative: it has a much higher $T_{\mathrm{c}}\geq$ 6.2 K and no magnetic order. Here we report a hallmark signature of spin-triplet superconductivity arising from the internal spin degrees of freedom via nuclear magnetic resonance measurements, and demonstrate the high tunability of the topological phases. We discovered three distinct superconducting phases and revealed the evolution of the paired-spins direction ($\mathbf{d}(\mathbf{k})$-vector). At low magnetic fields, $\mathrm{K_2Cr_3As_3}$ evolves from a helical (Phase A) to a chiral state (Phase B) with a rotation of the $\mathbf{d}(\mathbf{k})$-vector from in-plane to out-of-plane direction upon cooling, although both phases have point nodes in the gap. A line-nodal gap is realized in the high-field Phase C, where the $\mathbf{d}(\mathbf{k})$-vector lies in the basal plane. These findings establish $\mathrm{K_2Cr_3As_3}$ as a model spin-triplet superconductor and a promising platform for manipulating topological phases.
\end{abstract}

\maketitle

Two electrons with individual spin $s$ = 1/2 can form Cooper pairs with total spin $S$ = 0 (spin singlet) or $S$ = 1 (spin triplet) to carry electrical currents without dissipation. For spin-triplet pairing, because of internal degree of freedom arising from $S$ = 1, various novel phenomena such as multiple superconducting phases with different gap symmetry and paired-spin orientations, half quantized vortices, vestigial orders, and collective modes, are expected and most of which have indeed been seen in superfluid $\mathrm{^3}$He [1, 2]. However, progress in research on spin-triplet superconductivity has been slow, despite extensive efforts in doped topological insulator [3, 4] and certain U-based compounds [5]. This is largely due to the low critical temperature ($T_{\mathrm{c}}$) in these systems and the strong 5$f$-electron fluctuations in U-based compounds. These fluctuations, combined with coexisting and competing orders, often lead to contradictory conclusions and hinder the elucidation of intrinsic properties [6--8]. Exploring non-magnetic platforms with high $T_{\mathrm{c}}$ is essential for understanding the physics of the exotic pairing symmetry. This is particularly relevant for elucidating topologically non-trivial superconducting states, as spin-triplet superconductors with odd parity or broken time-reversal symmetry can harbor edge states---known as Majorana excitations---on surfaces or in vortex cores [9], which may be utilized for fault-tolerant quantum computing [10]. Furthermore, establishing methods to manipulate topological phases is a key step toward practical applications of topological superconductivity.

Recently, a new family of superconductors containing 3$d$ transition metal chromium, $A\mathrm{_2Cr_3As_3}$ ($A$ = Na, K, Rb, and Cs) with the transition temperature $T_{\mathrm{c}}$ up to 8.6 K, has been discovered [11--14]. The key building unit of $\mathrm{K_2Cr_3As_3}$ is a double-walled [($\mathrm{Cr_3As_3}$)$^{2-}$]$_{\infty}$ sub-nanotube elongated along the $c$-axis (inset of Fig. 1). The inner wall of this tube consists of a chromium triangle piece, while the outer wall is composed of arsenic. The crystal structure belongs to point group $D_{3h}$ (space group $P\bar{6}m2$). Notably, the upper critical field, $B_{\mathrm{c2}}$, far exceeds the Pauli paramagnetic limit in both the $\mathbf{B}$$\parallel$$c$ and $\mathbf{B}$$\perp$$c$ directions [15]. The spin-lattice relaxation rate 1/$T\mathrm{_1}$ at zero magnetic field exhibits a $T^{5}$ dependence without a coherence peak just below $T_{\mathrm{c}}$, indicating point nodes in its superconducting gap function [16]. The Knight shift measured by nuclear magnetic resonance (NMR), which is proportional to the local spin susceptibility  $\chi_{\mathrm{spin}}$, shows an anisotropic response; it remains constant under $\mathbf{B}$$\perp$$c$ but vanishes under $\mathbf{B}$$\parallel$$c$ in the superconducting state, indicating a spin-triplet superconducting state [17]. Recently, phase-sensitive measurements have obtained evidence for sign change of the wave function [18].  There are three Fermi surfaces, with the three-dimensional one making a dominant contribution to the density of states (DOS) [19], and the one-dimensional ones seem to contribute little to superconductivity [20]. Thus the unconventional superconducting state appears to arise from the three-dimensional Fermi surface. Although nuclear quadrupole resonance (NQR) and NMR measurements revealed the development of ferromagnetic spin fluctuations [16, 17, 21], no magnetic order has been found. In this way, $\mathrm{K_2Cr_3As_3}$, with its high $T_{\mathrm{c}}$ and the absence of long-range magnetic order that could interfere with superconductivity, offers an ideal platform for investigating the intrinsic properties of spin-triplet superconductivity and a platform to explore the manipulation of topological phases. Identifying multiple superconducting phases is another route to establishing spin-triplet pairing, as the internal spin degrees of freedom allow for various $\mathbf{d}(\mathbf{k})$-vector configurations. Analogous to superfluid $\mathrm{^3}$He, such phases are expected to emerge as a function of temperature or magnetic field. 

\begin{figure}[t]  
	\centering
	\includegraphics[width=0.95\columnwidth]{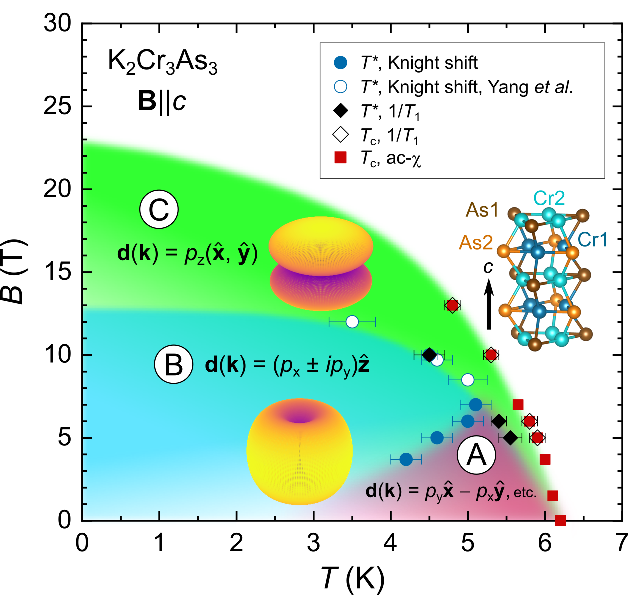}
	\caption{Multiple superconducting phases of $\mathrm{K_2Cr_3As_3}$ for $\mathbf{B}$$\parallel$$c$. The $T_{\mathrm{c}}$ values, represented by red squares and open diamonds, were determined by ac susceptibility and 1/$T\mathrm{_1}$ measurements. The characteristic temperatures $T^{\ast}$, represented by blue circles (open and filled) and black diamonds, were determined from the Knight shift and/or 1/$T\mathrm{_1}$ measurements. The errors in $T^{\ast}$ represented by blue circles were estimated by assuming that the uncertainty is equal to the point (temperature) interval around the position indicated by the dashed arrow in Fig. 2(a).  The errors in $T^{\ast}$ represented by black diamonds were estimated from the 1/$T\mathrm{_1}$ measurement temperature interval. The illustrations within Phases A/B and C display the superconducting gap functions with a linear combination of $p_x$ and $p_y$ [$\Delta$($\theta$, $\phi$) = $\Delta_{\mathrm{0}}$$\sin\theta e^{i\phi}$] and for the $p_z$-wave [$\Delta$($\theta$, $\phi$) = $\Delta_{\mathrm{0}}$$\cos\theta e^{i\phi}$], respectively. See also Fig. 6 in the End Matter for the spin structure of each phase.}
	\label{fig:top_image}
\end{figure}

In this Letter, we investigate the superconducting state of $\mathrm{K_2Cr_3As_3}$ by $\mathrm{^{75}}$As-NMR measurements over a wide range of magnetic field and temperature. We identified three distinct superconducting phases characterized by different $\mathbf{d}(\mathbf{k})$-vector configurations for $\mathbf{B}$$\parallel$$c$. Furthermore, we show that these phases possess different topological properties that can be tuned by external parameters. Our results provide new compelling evidence for spin-triplet superconductivity in this system. The results are summarized in Fig. 1 and Fig. 6 in the End Matter, and we describe the details below.

To explore the evolution of the paired spins ($\mathbf{d}(\mathbf{k})$-vector) direction, we conducted $\mathrm{^{75}}$As-NMR Knight shift $K$ measurements for the As2 site at magnetic fields of $B$$_{\parallel}$$_c$ = 3.7, 5.0, 6.0, 7.0, and 10 T. All experiments were performed using single crystals grown by the high-temperature solution growth method [22, 23]. $K$ is a sensitive probe of $\chi_{\mathrm{spin}}$, and is a powerful tool for discerning spin symmetry in the superconducting state. In a spin-singlet state, $\chi_{\mathrm{spin}}$ vanishes below $T_{\mathrm{c}}$, leading to a concomitant decrease in $K$ for all magnetic field directions. Conversely, in a spin-triplet state, $K$ decreases below $T_{\mathrm{c}}$ only when the $\mathbf{d}(\mathbf{k})$-vector possesses a component parallel to the magnetic field. Figure 2(a) shows the temperature and field dependences of $K$. We first discuss the result for $B$$_{\parallel}$$_c$ = 10 T which is shown in the top panel of Fig. 2(a). The $K$ does not decrease from $T_{\mathrm{c}}$ = 5.3 K determined by in-situ ac susceptibility (Fig. S1 in the Supplemental Material [23]) and 1/$T\mathrm{_1}$ measurements [Fig. 3(b)], but rather decreases at a lower temperature $T^{\ast}$ = 4.5 K. The raw data, namely, the temperature dependence of the NMR spectra at $B$$_{\parallel}$$_c$   = 10 T is shown in Fig. 2(b). The peak of the spectrum remains unchanged between $T_{\mathrm{c}}$ and $T^{\ast}$ but shifts clearly to a lower frequency below $T^{\ast}$. Similar temperature dependence was observed at $B$$_{\parallel}$$_c$ = 7.0 T, exhibiting a behavior consistent with previous data [21] that $T^{\ast}$ increases as the magnetic field decreases. This result means that above these magnetic fields, the $\mathbf{d}(\mathbf{k})$-vector is perpendicular to the $c$-axis but is parallel to the $c$-axis below these field strengths. This allows one to draw a phase boundary between Phases B and C. 

However, at magnetic fields below 7 T, the Knight shift shows an unexpected temperature dependence. As shown in Fig. 2(a) for $B$$_{\parallel}$$_c$  = 6.0, 5.0, and 3.7 T, $T^{\ast}$ decreases as the magnetic field is reduced, resulting in a distinct kink in the phase boundary below 7 T (filled blue circles in Fig. 1). As will become clear later, the filled blue circles in Fig. 1 represent a boundary between Phase B and a new Phase A across which the $\mathbf{d}(\mathbf{k})$-vector rotates by 90 degrees too.

The key question is whether the change in the $\mathbf{d}(\mathbf{k})$-vector direction is simply due to an SO(3) spin rotation or is due to a transition accompanying a change in the gap function. To address this issue, we measured the temperature dependence of 1/$T\mathrm{_1}$. It is widely recognized that 1/$T\mathrm{_1}$ is sensitive to the gap function through the DOS, $N$$\mathrm{_s}$($E$), in the superconducting state as 
\begin{equation*}
	\frac{T_{1}^{-1}(T)}{T_{1}^{-1}(T_{\mathrm{c}})} = \frac{2}{k_{\mathrm{B}} T} \int \left( \frac{N_{\mathrm{s}}(E)}{N_{0}} \right)^{2} f(E) [1 - f(E)] dE,
\end{equation*}
where $N\mathrm{_s}(E)/N_{0}$ = $E/\sqrt{(E^2 - \Delta^2)}$ with $N$$_{0}$ being the DOS in the normal state and $f(E)$ is the Fermi distribution function. For conventional BCS superconductors, 1/$T\mathrm{_1}$ will show a coherence peak just below $T_{\mathrm{c}}$ and decreases following an exponential $T$-dependence, 1/$T\mathrm{_1}$ $\propto$  exp($-\Delta/k_{\mathrm{B}}T$). For a line-nodal gap, $N$$\mathrm{_s}$($E$) $\propto$ $E$, and thus 1/$T\mathrm{_1}$ $\propto$  $T^3$ as seen in high-$T_{\mathrm{c}}$ cuprates [24] and many heavy-fermion compounds [25]. While for a gap with point nodes, $N$$\mathrm{_s}$($E$) $\propto$  $E\mathrm{^2}$, so that 1/$T\mathrm{_1}$ $\propto$  $T\mathrm{^5}$.
Figure 3 shows the temperature dependence of 1/$T\mathrm{_1}$ at $B$$_{\parallel}$$_c$  = 6.0, 10 and 13 T. At $B$$_{\parallel}$$_c$  = 13 T, 1/$T\mathrm{_1}$ decreases steeply just below $T_{\mathrm{c}}$ due to the superconducting transition and shows no coherence peak. Furthermore, 1/$T\mathrm{_1}$ is proportional to $T\mathrm{^3}$ below $T \sim 0.5T_{\mathrm{c}}$. This temperature dependence indicates the presence of line nodes in the gap function in Phase C. At $B$$_{\parallel}$$_c$  = 10 T and 6.0 T, however, 1/$T\mathrm{_1}$ shows a different temperature dependence; although 1/$T\mathrm{_1}$ decreases below $T_{\mathrm{c}}$ at both fields, it shows a distinct change in slope at $T$ = 4.5 K (10 T) and 5.4 K (6 T), obeying a $T\mathrm{^5}$ variation, indicative of point nodes in the superconducting gap, similar to that observed in the NQR measurement at zero magnetic field [16]. The temperature at which 1/$T\mathrm{_1}$ changes slope is defined as $T^{\ast}$. Figure 4 in the End Matter shows the normalized 1/$T\mathrm{_1}$ data for $B$$_{\parallel}$$_c$  = 13, 10, 6.0 and 5.0 T. 

\begin{figure}[t]  
	\centering
	\includegraphics[width=\columnwidth]{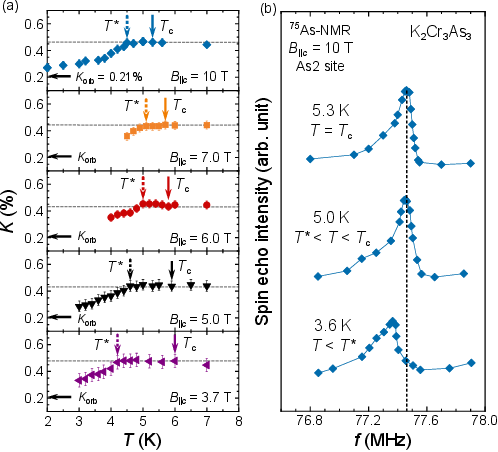}
	\caption{Temperature dependence of Knight shift and $^{75}$As-NMR spectra. (a) Temperature and field dependences of $K$ for $\mathbf{B}$$\parallel$$c$. The solid arrow indicates $T_{\mathrm{c}}$. The dashed arrow indicates $T^{\ast}$, the temperature at which $K$ begins to decrease upon further cooling. The error in $K$ was estimated by assuming that the peak uncertainty equaled the frequency resolution of the NMR data. The value of $K_{\mathrm{orb}} = 0.21\%$ was determined from the linear relationship between $K$ and bulk susceptibility $\chi$ [22]. (b) Temperature dependence of $^{75}$As-NMR spectra (the center peak; $+1/2 \leftrightarrow -1/2$ transition) at $B_{\parallel c} = 10$~T for the As2 site. At $B_{\parallel c} = 10$~T, $T_{\mathrm{c}}$ is 5.3~K, and $T^{\ast}$ is 4.5~K. The dashed line indicates the peak position of the NMR spectrum at $T_{\mathrm{c}}$.}
\label{fig:top_image}
\end{figure}

The change in the slope of 1/$T\mathrm{_1}$ observed at $B$$_{\parallel}$$_c$  = 10, 6.0, and 5.0 T is seen more clearly in the plot of the quantity 1/$T\mathrm{_1}$$T$ versus $T$, as shown in Fig. 5 in the End Matter. At $B$$_{\parallel}$$_c$  = 13 T, 1/$T\mathrm{_1}$$T$ exhibits a sharp decrease upon entering the superconducting state. At $B$$_{\parallel}$$_c$  =10 T, however, 1/$T\mathrm{_1}$$T$ initially decreases in a similar manner to that at 13 T but shows a more rapid decrease below $T^{\ast}$= 4.5 K. As the magnetic field is lowered to 6.0 T and 5.0 T, the difference between $T_{\mathrm{c}}$ and $T^{\ast}$ becomes narrower. 

As described above, the transition from $T^3$ to $T^5$ behavior in 1/$T\mathrm{_1}$ characterizes a fundamental change in the gap symmetry at $T^{\ast}$. Consequently, the boundary marked by $T^{\ast}$ (black diamonds in Fig. 1) separates the line-nodal gap state and point-nodal gap state. At $B$$_{\parallel}$$_c$ = 10 T, $T^{\ast}$ determined by the Knight shift and 1/$T\mathrm{_1}$ is found to coincide, indicating that both the Knight shift and 1/$T\mathrm{_1}$ probe the same transition between Phase B and Phase C. Thus, in this magnetic field region, the reorientation of the $\mathbf{d}(\mathbf{k})$-vector and the change in the superconducting gap symmetry occur simultaneously. Most interestingly, this phase boundary extends to higher temperature as the field is lowered to 6.0 T and 5.0 T. Namely, the phase boundary splits into two as the field is lowered to 6.0 T and 5.0 T. This result indicates that, below 7 T, the superconducting state undergoes two successive changes as the temperature is lowered—a change of the gap structure followed by a reorientation of the $\mathbf{d}(\mathbf{k})$-vector—thereby a new Phase A is revealed.

\begin{figure}[t]  
	\centering
	\includegraphics[width=\columnwidth]{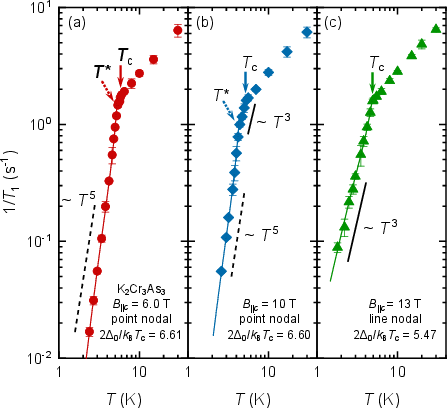}
	\caption{Temperature dependence of the $^{75}$As-NMR $1/T_1$ for $\mathbf{B}$$\parallel$$c$. (a)--(c) Temperature dependence of $1/T_1$ for $B_{\parallel c} = 6.0$, 10, and 13~T. Solid and dashed arrows indicate $T_{\mathrm{c}}$ and $T^{\ast}$, respectively. The red, blue, and green lines represent the results of fitting the data to a point-nodal gap with $2\Delta_0/k_{\mathrm{B}}T_{\mathrm{c}} = 6.61$ and 6.60, and a $p_z$-wave line-nodal gap with $2\Delta_0/k_{\mathrm{B}}T_{\mathrm{c}} = 5.47$, respectively. The $1/T_1$ data were fitted below $T^{\ast}$ at 6.0 and 10~T and $T_{\mathrm{c}}$ at 13~T. The dashed and solid black lines serve as guides for $T^5$ and $T^3$, respectively. Error bars represent the standard deviations in theoretical fitting of the nuclear magnetization recovery curves.}
	\label{fig:top_image}
\end{figure}

\begin{table*}[t] 
	\caption{Spin-triplet, odd-parity superconducting order parameters allowed in point group $D_{3h}$ and the expected physical properties.}
	\label{tab:summary}
	\centering
	\begin{tabular}{cccccc}
		\hline \hline
		$\Gamma$ & $\mathbf{d}(\mathbf{k})$-vector & Nodes & $\chi_{\mathrm{spin}} (\mathbf{B} \parallel c)$ & $1/T_1$ & Superconducting phase \\ \hline
		$A_1''$ & $p_x \hat{\mathbf{x}} + p_y \hat{\mathbf{y}}$ & Point ($k_z = 0$) & Remains constant & $\propto T^5$ & Phase A \\
		$A_2''$ & $p_y \hat{\mathbf{x}} - p_x \hat{\mathbf{y}}$ & Point ($k_z = 0$) & Remains constant & $\propto T^5$ & Phase A \\
		$E''$   & $p_x \hat{\mathbf{x}} - p_y \hat{\mathbf{y}}$, $p_y \hat{\mathbf{x}} + p_x \hat{\mathbf{y}}$ & Point ($k_z = 0$) & Remains constant & $\propto T^5$ & Phase A \\
		$E'$    & $(p_x \pm i p_y) \hat{\mathbf{z}}$ & Point ($k_z = 0$) & Decrease & $\propto T^5$ & Phase B \\
		$E'$    & $p_z (\hat{\mathbf{x}}, \hat{\mathbf{y}})$ & Line ($k_x$-$k_y$ plane) & Remains constant & $\propto T^3$ & Phase C \\ \hline \hline
	\end{tabular}
\end{table*}

We discuss the nature of each phase shown in Fig. 1. Table $\mathrm{I}$ lists the spin-triplet order parameters allowed for $D_{3h}$ point group [17]. The present measurements reveal two key findings. First, the gap in Phase C is found to be line nodal, as evidenced by a $T^3$ dependence of 1/$T\mathrm{_1}$. Given that the Knight shift shows no decrease below $T_{\mathrm{c}}$ (Fig. S4 in the Supplemental Material [23]), this is evidence for the realization of $\mathbf{d}(\mathbf{k}) = p_{z}(\hat{\mathbf{x}}, \hat{\mathbf{y}})$ belonging to the $E'$ representation. The gap function is illustrated as an inset to Fig. 1. Indeed, the $p_z$ gap function $\Delta$($\theta$, $\phi$) = $\Delta_{\mathrm{0}}$$\cos\theta e^{i\phi}$  well fits the 1/$T\mathrm{_1}$ results as shown in Fig. 3(c). This state was previously proposed theoretically for the superconducting phase at zero field [26--28]. Even for a simple $p_{x, y, z}$-wave superconducting state, it was shown that the winding number is non-zero and such states can also be topological [29]. Therefore, a 90-degrees rotation of the $\mathbf{d}(\mathbf{k})$-vector across the B-C boundary marks a transition to a totally different phase with a different gap structure and different topology (see below). 

The second key result is the finding of a new Phase A in the high-temperature, low-magnetic-field regime. In Phase A, in contrast to Phase B, the Knight shift does not decrease below $T_{\mathrm{c}}$. Also being different from Phase C, 1/$T\mathrm{_1}$ follows a $T^5$ dependence in this new phase. All the properties of Phase A are consistent with helical states belonging to the representations of $A_1''$ [$\mathbf{d}(\mathbf{k}) = p_{x}\hat{\mathbf{x}} + p_{y}\hat{\mathbf{y}}$], $A_2''$ [$\mathbf{d}(\mathbf{k}) = p_{y}\hat{\mathbf{x}} - p_{x}\hat{\mathbf{y}}$], or $E''$ [$\mathbf{d}(\mathbf{k}) = p_{x}\hat{\mathbf{x}} - p_{y}\hat{\mathbf{y}},  p_{y}\hat{\mathbf{x}} + p_{x}\hat{\mathbf{y}}$] (Table $\mathrm{I}$). There are point nodes along the $k_z$ direction in the gap function of these states, but the $\mathbf{d}(\mathbf{k})$-vector lies in the $ab$-plane so that the parallel spins of the Cooper pairs are along the $c$-axis (Fig. 6 in the End Matter). As seen in Fig. 3(a), a point nodal gap $\Delta$($\theta$, $\phi$) = $\Delta_{\mathrm{0}}$$\sin\theta e^{i\phi}$ can well fit the data. Such states can be stabilized due to a gain of Zeeman energy when a magnetic field is applied along the $c$-axis. It is shown that nematic interactions, such as local strain, could  also promote the emergence of helical states from a chiral state [30]. It is pointed out that the helical states, in particular the $A_1''$ state, are similar to a two-dimensional projection of the Balian-Werthamer state [31] realized in $^3$He-B phase [1]. It is also important to note that the helical states are topological. In particular, a thin film with a thickness smaller than the superconducting coherence length along the $c$-axis can host gapless counter-propagating Majorana states at the boundary [32]. 

The transition from Phase B to Phase C can be understood as due to a Zeeman energy gain. The spins of the Cooper pairs are oriented within the basal plane in Phase B. Upon increasing the magnetic field along the $c$-axis, Zeeman energy gain can lead to a transition from B to C where the spins have components along the $c$-axis. Consequently, the phase boundary between Phases B and C is determined by the competition between the pinning force acting on the $\mathbf{d}(\mathbf{k})$-vector and the Zeeman energy. In fact, the low-temperature phase boundary extrapolated from the high-temperature 
agrees with the Pauli-limited critical field $B\mathrm{_{c2}^P}\sim\mathrm{13~T}$. It is emphasized that Phase B breaks time reversal symmetry as supported by muon spin rotation relaxation measurement [33] and is a three-dimensional analogue of the superfluid $^3$He-A phase [1] described by the so-called ABM model [34, 35]. The vortex cores can host Majorana excitations [9].

In summary, our results establish $\mathrm{K_2Cr_3As_3}$ as a rare platform for probing the intrinsic nature of spin-triplet superconductivity, and by extension, for studying topological superconductivity. We identified three different superconducting phases for $\mathbf{B}$$\parallel$$c$ and elucidated all the structure of the paired spins and gap structures. It would be an interesting future task to explore the phase diagram for $\mathbf{B}$$\perp$$c$ to which the NQR 1/$T\mathrm{_1}$ results may belong, as the quantized axis is in the plane for that measurement.

The identified three distinct superconducting phases and the associated evolution of the $\mathbf{d}(\mathbf{k})$-vector relative to the temperature and magnetic field would be inexplicable within a spin-singlet framework. Rather, this behavior is a hallmark of spin-triplet superconductivity with an internal spin degree of freedom, providing new compelling evidence for the realization of spin-triplet pairing in $\mathrm{K_2Cr_3As_3}$. At low magnetic fields and high temperatures close to $T_{\mathrm{c}}$, helical state(s) (Phase A) with point nodes in the gap function emerges, which evolves into a chiral, Weyl state upon cooling to lower temperatures (Phase B). Our Phase A (in particular, the $A_1''$ state) is similar to the 2D projection of the superfluid $^3$He-B phase, while our Phase B is analogous to the superfluid $^3$He-A phase [1]. Instead of pressure as a tuning external parameter in superfluid $^3$He, it is the magnetic field in $\mathrm{K_2Cr_3As_3}$. Furthermore, Phases A and B are superconducting analogs to Weyl semimetals or topological Skyrmion magnets [36, 37], in which anomalous Hall effect (AHE) has been observed. The high-field phase C has a line nodal gap which shares similarities with many unconventional superconductors including the high-$T_{\mathrm{c}}$ copper oxides and many heavy-fermion compounds.

Thus, $\mathrm{K_2Cr_3As_3}$ is a prototypical three-dimensional spin-triplet superconductor which provides a platform to study topological phenomena in general and, most importantly, to lay a material foundation for realizing topological quantum computing. In this work, we demonstrated the high tunability of various topological phases by magnetic field and temperature. Looking ahead, it would also be a fertile frontier to explore the manipulation of the various phases via a ferromagnetic junction on the surface of $\mathrm{K_2Cr_3As_3}$.

$Acknowledgements$- We thank Kazushige Machida, Yoichi Yanase, Mario Cuoco, Yukio Tanaka, Masatoshi Sato, Fei Zhou, Yi Zhou and Jie Yang for stimulating and helpful discussions. This work was supported by JSPS KAKENHI Grant Numbers 19H00657, 22H04482, 26K07015, and the Okayama Foundation for Science and Technology, 2024.

$Data~availability$- The data used in this work are not publicly available but are available from the authors upon reasonable request. Correspondence and requests for materials should be addressed to G.-Q.Z..

\clearpage
\begin{center}
	\bfseries\MakeUppercase{End Matter}
\end{center}

\textit{Superconducting gap function}---Here, we discuss the superconducting gap structure inferred from the present work. To begin our analysis of the superconducting gap function, we define the primitive translation vectors of the hexagonal lattice as $\mathbf{a} = (a/2, -\sqrt{3}a/2, 0)$, $\mathbf{b} = (a/2, \sqrt{3}a/2, 0)$, $\mathbf{c} = (0, 0, c)$. The corresponding inner products with the crystal momentum $\mathbf{k}$ are given by $k_a = \mathbf{k} \cdot \mathbf{a}$, $k_b = \mathbf{k} \cdot \mathbf{b}$, $k_c = \mathbf{k} \cdot \mathbf{c}$, respectively. Noting that the lattice constant along the $a$-axis in $\text{K}_2\text{Cr}_3\text{As}_3$ is more than twice that along the $c$-axis ($a = 9.9832$~\AA, $c = 4.2304$~\AA) [11], we consider nearest-neighbor hopping within the $ab$-plane and second-nearest-neighbor hopping along the $c$-axis. Under these constraints, the $p$-wave harmonic functions take the following form:
\begin{align*}
	&p_x(\mathbf{k})\propto2\sin(k_a+k_b)+\sin(k_a)+\sin(k_b)\\
	&p_y(\mathbf{k})\propto\sqrt{3}\sin(k_a-k_b)\\
	&p_z(\mathbf{k})\propto\sin(k_c)+2\sin(2k_c)
\end{align*}
Furthermore, in the long-wavelength limit near the Fermi energy, where $k_a, k_b, k_c$ $<$$<$ 1, the harmonic functions can be approximated as:
\begin{align*}
	&p_x(\mathbf{k})\propto k_x \propto \sin\theta\cos\phi\\
	&p_y(\mathbf{k})\propto k_y \propto \sin\theta\sin\phi\\
	&p_z(\mathbf{k})\propto k_z \propto \cos\theta
\end{align*}
Here, $\theta$  and $\phi$  represent the polar and azimuthal angles, respectively. 
First, we discuss the superconducting gap structure in Phase A with three possible helical candidates that give rise to the same NMR results. Taking $\mathbf{d}(\mathbf{k}) = p_{x}(\hat{\mathbf{x}}) + p_{y}(\hat{\mathbf{y}})$ state ($A_1''$ state) as a representative example. $\mathbf{d}(\mathbf{k})$-vector in the $A_1''$ state can be represented by:
\begin{align*}
	\mathbf{d}(\mathbf{k}) \propto \begin{pmatrix} k_x \\ k_y \\ 0 \end{pmatrix}
\end{align*}
The magnitude of the $\mathbf{d}(\mathbf{k})$-vector is equal to that of the superconducting gap ($|\mathbf{d}(\mathbf{k})|=|\Delta(\mathbf{k})|$). Therefore, $|\Delta(\mathbf{k})|\propto\sqrt{k_x^2+k_y^2}$, and $|\Delta(\mathbf{\theta})|$ is obtained as:
\begin{align*}
	|\Delta(\mathbf{\theta})| \propto \sqrt{\sin^2\theta(\sin^2\phi+\cos^2\phi)}=\sin\theta
\end{align*}
The other two candidates, namely the $A_2''$ and $E''$ states have the same gap structure as $A_1''$ state.

Next, we discuss the superconducting gap structure in Phase B. The order parameter of Phase B suggested by NMR/NQR experiments [16, 17] is $\mathbf{d}(\mathbf{k}) = (p_{x}\pm i p_{y})\hat{\mathbf{z}}$. Therefore, $\mathbf{d}(\mathbf{k})$-vector in Phase B can be expressed as:
\begin{align*}
	\mathbf{d}(\mathbf{k}) \propto \begin{pmatrix} 0 \\ 0 \\ k_x\pm ik_y \end{pmatrix}
\end{align*}
Hence, $|\Delta(\mathbf{k})|\propto\sqrt{k_x^2+k_y^2}$, which leads to $|\Delta(\mathbf{\theta})|$ expressed as:
\begin{align*}
	|\Delta(\mathbf{\theta})| \propto\sin\theta
\end{align*}
This is precisely the so-called ABM gap with point nodes along $k_z$ = 0 [34]. 

Finally, in Phase C, the order parameter $\mathbf{d}(\mathbf{k})$-vector suggested by present work is $\mathbf{d}(\mathbf{k}) = p_{z}(\hat{\mathbf{x}}, \hat{\mathbf{y}})$. For the $\mathbf{d}(\mathbf{k}) = p_{z}\hat{\mathbf{x}}$ state, it can be written as:
\begin{align*}
	\mathbf{d}(\mathbf{k}) \propto \begin{pmatrix} k_z \\ 0 \\ 0 \end{pmatrix}
\end{align*}
Therefore,  $|\Delta(\mathbf{k})|\propto\sqrt{k_z^2}$. Consequently, $|\Delta(\mathbf{\theta})|$ can be expressed as:
\begin{align*}
	|\Delta(\mathbf{\theta})| \propto\cos\theta
\end{align*}
This is precisely the polar gap with a line node in the $k_x$-$k_y$ plane [35], and same results are obtained for any $\mathbf{d}(\mathbf{k}) = p_{z}(\hat{\mathbf{x}}, \hat{\mathbf{y}})$ state.
Figure 6 is an illustration of the vector order parameter $\mathbf{d}(\mathbf{k})$-vector and the paired parallel spins.

\begin{figure}[H]
	\centering
	\includegraphics[width=\columnwidth]{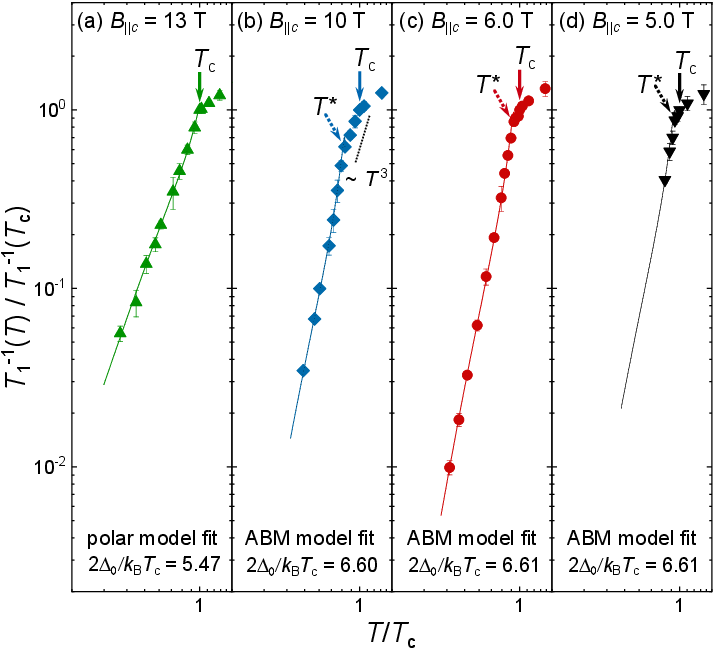}
	\caption{Temperature dependence of the normalized $^{75}$As-NMR 1/$T_1$ for $\mathbf{B}$$\parallel$$c$. (a)--(d) Temperature dependences of the normalized 1/$T_1$ for $\mathbf{B}$$\parallel$$c$ at 13, 10, 6.0, 5.0 T are shown. Solid and dashed arrows indicate $T_{\mathrm{c}}$ and $T^{\ast}$, respectively. The green, blue, red, and black lines represent the results of fitting the polar model ($2\Delta_0/k_{\mathrm{B}}T_{\mathrm{c}} = 5.47$) and the ABM model ($2\Delta_0/k_{\mathrm{B}}T_{\mathrm{c}} = 6.60, 6.61$ and 6.61) to the data. The polar model fitting was performed below $T_{\mathrm{c}}$, and the ABM model fitting was performed below $T^{\ast}$. Error bars represent the standard deviations of the parameters. The dashed line serves as a guide for the temperature variation of $T^3$. }.
\end{figure}
\begin{figure}[H]
	\centering
	\includegraphics[width=0.772\columnwidth]{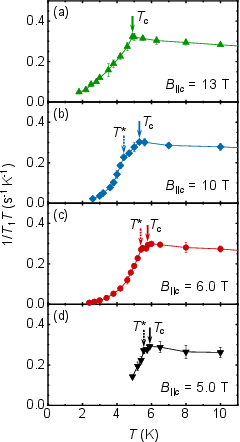}
	\caption{Temperature dependence of $^{75}$As-NMR 1/$T_1T$ for $\mathbf{B}$$\parallel$$c$. (a)--(d) Temperature and field dependences of 1/$T_1T$ for $\mathbf{B}$$\parallel$$c$ at 13, 10, 6.0, 5.0 T. Solid and dashed arrows indicate $T_{\mathrm{c}}$ and $T^{\ast}$, respectively. Error bars represent the standard deviations in theoretical fitting the nuclear magnetization recovery curves.}.
\end{figure}

\raggedbottom
\textit{The role of antisymmetric spin-orbit coupling (ASOC) arising from inversion symmetry breaking}---$\mathrm{K_2Cr_3As_3}$ crystallizes in the non-centrosymmetric point group $D_{3h}$, allowing for a finite ASOC characterized by the $\mathbf{g}(\mathbf{k})$-vector. Generally, the $\mathbf{d}(\mathbf{k})$-vector tends to align parallel to the $\mathbf{g}(\mathbf{k})$-vector to minimize energy [38]. The $\mathbf{g}(\mathbf{k})$-vector for $D_{3h}$ symmetry is given by:
$\mathbf{g}(\mathbf{k}) =
\alpha_1 [2 k_x k_y k_z s_x - (k_x^2 - k_y^2)k_z s_y]
+ \alpha_2 (k_y^3 - 3 k_x^2 k_y) s_z$.
Unlike layered systems where quasi-two-dimensionality leads to strong pinning along specific directions, the superconductivity in $\mathrm{K_2Cr_3As_3}$ is dominated by the three-dimensional $\gamma$ band. This three-dimensionality is expected to effectively average out the anisotropy of the $\mathbf{g}(\mathbf{k})$-vector over the Fermi surface, thereby suppressing the pinning of the $\mathbf{d}(\mathbf{k})$-vector. This interpretation is supported by the very small magnetic anisotropy observed in the normal state [22] and ARPES measurements [39]. We propose that this weak ASOC competes with the Zeeman energy and orbital effects on comparable energy scales. Consequently, the $\mathbf{d}(\mathbf{k})$-vector orientation is determined by a delicate balance among these factors, making it highly sensitive to temperature and magnetic field. Such behavior, reminiscent of superfluid $^3$He but realized in a solid-state system, is of significant physical interest. Furthermore, the small ASOC and the resulting isotropic spin susceptibility in the normal state make $\mathrm{K_2Cr_3As_3}$ a distinct class of spin-triplet superconductors, setting it apart from many uranium-based candidates that exhibit strong Ising-like magnetic anisotropy [40--43].

\begin{figure}[t]
	\centering
	\includegraphics[width=0.95\columnwidth]{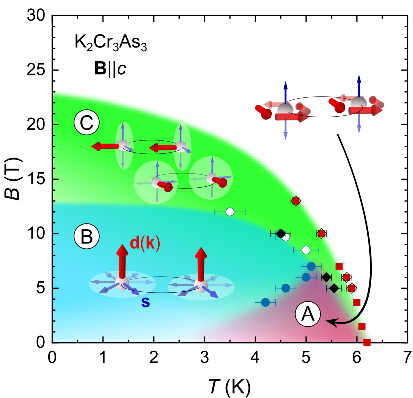}
	\caption{ Directions of the paired spins and $\mathbf{d}(\mathbf{k})$-vector in each phase. Schematic illustration of the Cooper pairs and $\mathbf{d}(\mathbf{k})$-vector in Phases A, B, and C. The white spheres represent electrons forming Cooper pairs. The Cooper pair spins $\mathbf{s}$  and the $\mathbf{d}(\mathbf{k})$-vector are indicated by blue and red arrows, respectively. A representative order parameter for Phase A, $\mathbf{d}(\mathbf{k}) = p_{y}\hat{\mathbf{x}} - p_{x}\hat{\mathbf{y}}$, is shown in the upper-right. This state can be regarded as a quantum superposition of the $S_{\mathrm{z}}$ = 1 and $S_{\mathrm{z}}$ = $-$1 components. For other two states of Phase A, although the $\mathbf{d}(\mathbf{k})$-vector structures differ, they all realize a superposition of $S_{\mathrm{z}}$ = $\pm$1 states. In Phase B, the $\mathbf{d}(\mathbf{k})$-vector is aligned along the $c$-axis, leading to an $S_{\mathrm{z}}$ = 0 configuration where the electron spins lie within the basal plane. In Phase C, the $\mathbf{d}(\mathbf{k})$-vector is oriented along the $a$-axis ($b^{\ast}$-axis), and the paired (parallel) spins lie within the $b^{\ast}c$-plane ($ac$-plane). In Phases B and C, paired spins within the white circles all give rise to the respective $\mathbf{d}(\mathbf{k})$-vector indicated. }.
\end{figure}

\vfill
\pagebreak

\end{document}